\documentclass[nanomaterials,article,accept,pdftex,moreauthors]{Definitions/mdpi}
\firstpage{1} 
\pubvolume{1}
\issuenum{1}
\articlenumber{0}
\pubyear{2023}
\copyrightyear{2022}
\externaleditor{Academic Editor: {Zhan'ao Tan}
}
\datereceived{3 April 2023} 
\daterevised{24 April 2023}
\dateaccepted{25 April 2023} 
\datepublished{} 
\hreflink{https://doi.org/} 

\usepackage[normalem]{ulem}
\usepackage{bm}

\Title{Reconstruction-Induced $\varphi_0$ Josephson Effect in Quantum Spin Hall Constrictions}

\TitleCitation{Reconstruction-Induced $\varphi_0$ Josephson Effect in Quantum Spin Hall Constrictions}

\Author{Lucia Vigliotti $^{1,}$*\orcidA{}, Fabio Cavaliere $^{1,2}$*\orcidD{}, Giacomo Passetti $^{3}$\orcidB{}, Maura Sassetti $^{1,2}$ and Niccolò Traverso Ziani $^{1,2}$\orcidC{}}

\AuthorNames{Lucia Vigliotti, Fabio Cavaliere, Giacomo Passetti, Maura Sassetti and Niccolò Traverso Ziani}

\AuthorCitation{Vigliotti, L.; Cavaliere, F.; Passetti, G.; Sassetti, M.; Traverso Ziani, N.}

\address{%
$^{1}$ \quad Dipartimento di Fisica, Università degli Studi di Genova, Via Dodecaneso 33, 16146 Genova, Italy;  sassetti@fisica.unige.it (M.S.); traversoziani@fisica.unige.it (N.T.Z.)\\
$^{2}$ \quad CNR-SPIN, Via Dodecaneso 33, 16146 Genova, Italy\\
$^{3}$ \quad Institut f\"ur Theorie der Statistischen Physik, RWTH Aachen University and JARA-Fundamentals of Future Information Technology, 52056 Aachen, Germany; passetti@physik.rwth-aachen.de}

\corres{Correspondence: lucia.vigliotti@edu.unige.it (L.V.); fabio.cavaliere@unige.it (F.C.)}

\abstract{The simultaneous breaking of time-reversal and inversion symmetry, in connection to superconductivity, leads to transport properties with disrupting scientific and technological potential. Indeed, the anomalous Josephson effect and the superconducting-diode effect hold promises to enlarge the technological applications of superconductors and nanostructures in general. In this context, the system we theoretically analyze is a Josephson junction (JJ) with coupled reconstructed topological channels as a link; such channels are at the edges of a two-dimensional topological insulator (2DTI). We find a robust $\varphi_0$ Josephson effect without requiring the presence of external magnetic fields. Our results, which rely on a fully analytical analysis, are substantiated by means of symmetry arguments: Our system breaks both time-reversal symmetry and inversion symmetry. Moreover, the anomalous current increases as a function of temperature. We interpret this surprising temperature dependence by means of simple qualitative arguments based on Fermi's golden rule.}

\keyword{Josephson junctions; topological insulators; symmetry breaking; $\varphi_0$ effect}

\begin{document}


\section{Introduction} \label{sec:introduction}
Superconductivity \cite{tinkham} is at the heart of many quantum technological applications \cite{barone}. For example, SQUIDS \cite{tinkham} and nano-SQUIDS \cite{nanosq} represent extremely useful sensors for magnetic fields, hence finding application in the most diverse settings. Additionally, QuBits based on superconductors \cite{supq} are prominent in quantum computation, a field gaining more and more relevance as quantum supremacy appears to be a concrete goal. Superconductive correlations are crucial for the generation of Majorana zero modes \cite{kitaev,maj2,prb} and \mbox{parafermions \cite{para,para2},} which could alleviate the demanding needs of error correction: Indeed, such topological quasiparticles are characterized by non-Abelian exchange statistics---which is expected to open the way to topologically protected quantum computation protocols---although it is fair to remark that such non-Abelian statistics have not been experimentally detected so far. \mbox{A further} renowned application of superconductivity is superconducting spintronics \cite{sst}. The aim of this sub-field of spintronics is to merge the advantages of spintronics on standard electronics, with the properties of Cooper pairs (CPs). Furthermore, superconductors find room in caloritronics \cite{calo}. The list of applications is, however, continuously updating.

Recently, superconducting structures with broken time-reversal and inversion symmetry \cite{symmetry0,symmetry,symmetry2} have been shown to allow for two connected and promising behaviors: the $\varphi_0$ Josephson effect \cite{shukrinov,phizero,phizero1,phizero2,phizero3,phizero4,phizero5,phizero6,alidoust1,alidoust2} and the superconducting diode effect \cite{diode0,diode,diode1,alidoust3,alidoust4}. The first consists in the presence of a finite Josephson current, called an anomalous Josephson current, in the absence of a phase difference between the two superconductors and has recently been experimentally observed \cite{phizeroexp}. It is remarkable because it can be used to design phase batteries \cite{battery,battery2}, and to drive superconducting circuits \cite{circuits} and superconducting \mbox{memories \cite{memory}.} The second, experimentally observed as well \cite{diodeexp0,diodeexp1,diodeexp2}, inspires great perspectives, given the analogy to its semiconducting counterpart that basically opened the way to modern electronics \cite{diode2}. While the breaking of inversion symmetry can be provided by the geometry of the nanostructure \cite{referee,hou,seoane} or by the microscopic lattice \cite{inversion,noncentrosymm}, the breaking of time-reversal symmetry is usually, although not always \cite{filippozzi,nonmag1,nonmag2,diodeexp2}, driven externally by means of applied magnetic fields.

Many of these striking advances in the functionalities of superconducting structures are related to the superconducting proximity effect \cite{tinkham}, which consists in the induction of a superconducting pairing in non-superconducting materials put in close contact to a superconductor. This has remarkable significance, because CPs in the proximitized material do not need to have the same properties as in the proximitizing superconductor. As an implication, despite most known superconductors hosting zero-momentum spin-singlet CPs, very exotic superconducting states can be implemented in proximitized materials. In these regards, materials with strong spin--orbit coupling play a special role. Spin--orbit coupled quantum wires, for example, enabled the realization of spinless $p$-wave \mbox{superconductivity \cite{maj2}} with the purpose of realizing Majorana fermions. A second class of materials with strong spin--orbit coupling is that of topological insulators (TIs), for which the effects of proximization with superconductors are countless and extremely promising \cite{qi}. Within this vast literature, a major role is played by two-dimensional topological insulators (2DTIs), where a gapped two-dimensional bulk coexists with symmetry protected metallic edge channels at the boundaries of the structure. 2DTIs have been successfully proximitized, and intriguing effects, such as missing Shapiro steps, have been observed \cite{deacon}. 

In this Article, we further analyze the properties of proximitized 2DTIs. Inspired by recent developments in the field of non-proximitized TIs, we propose a novel type of Josephson junction (JJ). Its main building block is represented by a narrow constriction between the edge states characterizing the 2DTI. Such edge states are reconstructed, meaning that the metallic states characterized by different spin polarization are spatially separated. Notably, we analytically show that the system hosts a $\varphi_0$ Josephson effect in the absence of external magnetic fields. Moreover we show that, surprisingly, the effect is reinforced by a temperature increase in a range of parameters. Indeed, for low temperature and weak tunneling between the topological edge channels, the anomalous current has a small zero temperature contribution and a quadratic one. We interpret such temperature dependence on the basis of a simple perturbative argument.

The rest of the Article is structured as follows: In Section \ref{sec:methods}, we introduce the main ideas of our proposal by qualitatively describing the setup, its symmetries, and its working mechanisms. Subsequently, we provide a theoretical model for the system and its properties. In Section \ref{sec:results}, we provide a quantitative analysis of the emerging $\varphi_0$ effect. In Section \ref{sec:discussion}, we discuss and interpret our findings, even on the basis of perturbative expansions and simple arguments. Finally, in Section \ref{sec:conc}, we summarize and draw our conclusions.


\section{Methods} \label{sec:methods}
\begin{figure}[H]
	   \includegraphics[width=0.95\textwidth]{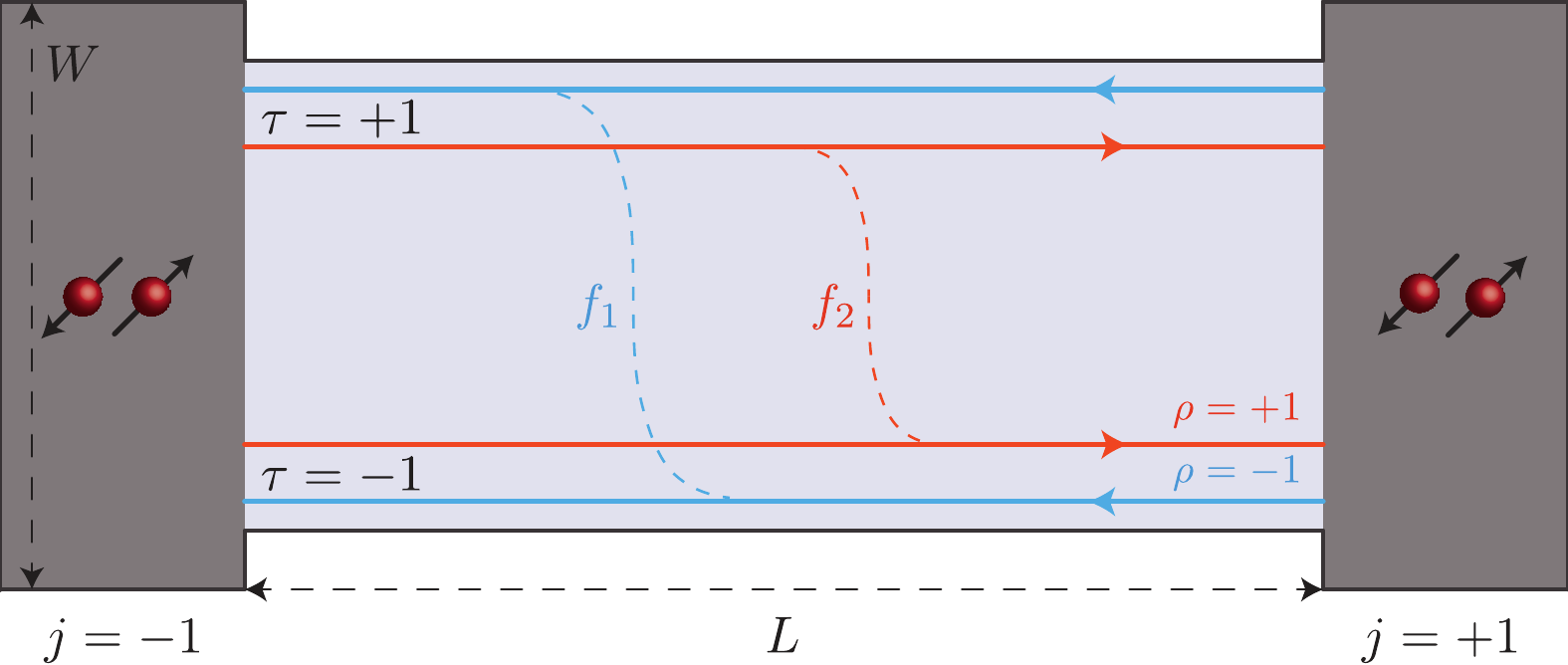}
	   \caption{Schematic of the system under consideration: a Josephson junction (JJ) of length $L$ and width $W$ with the edge channels of a two-dimensional topological insulator (2DTI) as non-superconducting region. The superconductors are assumed of $s$-wave type. Here, $\rho=\pm1$ labels the right/left-mover metallic channels, $\tau=\pm1$ the upper/lower edge, and $j=\pm1$ the right/left superconductor. Lastly, $f_{1/2}$ denotes the amplitude of the inter-edge tunneling for right/left-moving electrons. Due to the unbalance between the tunneling amplitudes, both inversion symmetry and time-reversal symmetry are broken in the system.}
	   \label{fig:setup}
    \end{figure}

The setup we consider is shown in Figure \ref{fig:setup} and consists of a JJ with a 2DTI as a normal part \cite{deacon,nanomaterials}.

The chemical potential of the TI is tuned inside the bulk gap, such that the relevant degrees of freedom are the ones related to the one-dimensional edge channels \cite{bhz,tiexp,tinl}. Crucially, such channels are helical, {i.e.,} on each edge electrons with opposite momentum have opposite spin projection. With reference to the picture, on the upper edge, right (left) moving electrons have spin up (down) projection. On the lower edge, the situation is reversed, {i.e.,} electrons with spin down (up) move right (left). At this stage, inversion symmetry is already broken at the level of the single edges, but time-reversal symmetry is not \cite{ti}. To break the latter, we follow a road provided by the recent advancements in the creation of constrictions between topological edge channels \cite{nat1}. The mechanism, which has three conceptual steps, is the following:

\begin{enumerate}
    \item By means of etching, the topological edge channels are brought at a distance that is comparable with their localization length. Electrons can then tunnel between the upper and lower edge. This can happen in two ways \cite{twotunn1,twotunn2,dolcettocimento}: via spin-preserving backscattering and via spin-flipping forward scattering.
    \item The constriction is made long with respect to the inverse Fermi momentum. This step is, to some extent, unavoidable: Currently, constrictions between helical edge states have only been performed in thick HgTe quantum wells, where the Dirac point is hidden in the bulk valence band \cite{nat1}. The spin-preserving backscattering is hence irrelevant, and the only process remaining is the spin-flipping forward scattering \cite{nat1}, denoted by $f_{\nu}$ (with $\nu=1,2$) in Figure \ref{fig:setup}. Even alone, such process is in any case still time-reversal invariant.
    \item The third step is to induce edge reconstruction: As proposed in Ref. \cite{reconstruction}, if the potential confining the edge channels is soft, the spin up and spin down channels can separate in real space, thus creating an unbalance in the tunneling rates for spin up and spin down electrons (see Figure \ref{fig:setup}). Time-reversal symmetry is hence also broken, as well as inversion symmetry at the level of the full structure.
\end{enumerate}
\vspace{-6pt}

Thanks to these three considerations, the JJ with reconstructed helical edges as a link is hence expected to exhibit a $\varphi_0$ Josephson effect in the absence of external fields.

Such anomalous effect is in a way similar to the one previously discussed in \cite{dolcini}, where, however, a magnetic field is required. Moreover, in that case, the effect is mostly present when the weak link is a single edge, while it tends to vanish when both edges \mbox{are considered.} 

We now provide a quantitative analysis of what we have just described. To begin, we introduce the fermionic operators $\hat{\psi}_{\rho,\tau}(x)$ that annihilate an electron at position $x$ propagating in the $\rho$-direction channel of the $\tau$ edge \cite{spinsass,biggio}. We set $\rho=1$ $(-1)$ for the right (left) direction of motion and $\tau=1$ $(-1)$ for the upper (lower) edge.
Due to the spin-momentum locking, these two indices completely define the edge states, since the spin polarization is $\tau\rho$.
For notational convenience, we introduce the spinor $\bm{\hat{\psi}}(x)=\left(\hat{\psi}_{1,1}(x),\hat{\psi}_{-1,1}(x),\hat{\psi}_{-1,-1}(x),\hat{\psi}_{1,-1}(x)\right)^T$. The Hamiltonian can hence be recast in the quadratic form
\vspace{-2pt}
    \begin{equation}
        \hat{H}=\hat{H}_E^0+\delta \hat{H}_E.
    \end{equation}
Here, $\hat{H}_E^0$ describes the 2DTI and $\delta \hat{H}_E$ the coupling to the superconductors, that are considered of $s$-wave type. Explicitly, we have

    \begin{equation}\label{eqn:H_E^0}
	   \hat{H}_E^0=\int_{-L/2}^{L/2} dx\, \bm{\hat{\psi}}^{\dagger}(x)\hat{\mathcal{H}}_E^0\bm{\hat{\psi}}(x),
    \end{equation}
where $L$ is the separation between the superconductors and $\hat{\mathcal{H}}_E^0$ is the Hamiltonian density of the edge channels, made up of two contributions: $\hat{\mathcal{H}}_E^0=\hat{\mathcal{H}}_{kin.}+\hat{\mathcal{H}}_{f.s.}$, that are the kinetic term and the forward scattering term, respectively. In detail,

\vspace{-10pt}
	\begin{align}
	    \hat{\mathcal{H}}_{kin.} &= \hbar v_F(-i\hat{\partial}_x)\hat{\tau}_3\otimes\hat{\rho}_3,\label{eqn:kin}\\
        \hat{\mathcal{H}}_{f.s.} &=\frac{f_1}{2}(-\hat{\tau}_2\otimes\hat{\rho}_2+\hat{\tau}_1\otimes\hat{\rho}_1)+\frac{f_2}{2}(\hat{\tau}_2\otimes\hat{\rho}_2+\hat{\tau}_1\otimes\hat{\rho}_1)\notag\\
        &=\frac{-f_1+f_2}{2}\hat{\tau}_2\otimes\hat{\rho}_2+\frac{f_1+f_2}{2}\hat{\tau}_1\otimes\hat{\rho}_1.\label{eqn:fs}
	\end{align}
    
In the equations above, $\hat{\rho}_i$ and $\hat{\tau}_i$ (with $i=0,1,2,3$) represent the identity and the three Pauli matrices acting in the right/left-movers and the upper/lower edge space, respectively, and $v_F$ is the Fermi velocity. The couplings $f_1$ and $f_2$ parameterize the tunneling between the edges: when $f_1\neq f_2$ the edges are reconstructed and time-reversal symmetry is broken. Our model does not include magnetic impurities, that would introduce intra-edge backscattering, due to the fact that long and ballistic edges can now routinely be synthetized. Furthermore, we neglect the possible presence of charge puddles in the bulk of the 2DTI since there is little room for them in a narrow sample. To summarize, the non-superconducting part just discussed represents a pair of one-dimensional Dirac cones, in which the two right-moving branches are shifted with respect to the left-moving ones proportionally to $f_1-f_2$. Moreover, branches with fixed chirality do not have fixed spin projection anymore, due to the spin non-conserving tunneling.

The lateral proximitizing superconductors, assumed to have a large superconducting gap with respect to all the other energy scales, influence the system via the perturbative Hamiltonian $\delta\hat{H}_E$. Besides the large gap hypothesis, its derivation requires a weak coupling between the edges and the superconductors, and the specific model of the latters. \mbox{In our} case, it is the standard BCS one, since in experiments on 2DTIs niobium is the prime example of employed superconductor \cite{deacon}. What one obtains is that the coupling to the right and left superconductors ($j=\pm 1$) is given by \cite{loss,njp} 
\vspace{-3pt}
    \begin{equation}
	   \delta \hat{H}_E= \sum_{\zeta_1,\zeta_2,j}\left[\gamma_{\zeta_1,\zeta_2,j}\hat{\psi}_{\zeta_1}(x_j^-)\hat{\psi}_{\zeta_2}(x_j^+)+h.c.\right]=\sum_{\zeta_1,\zeta_2,j}\left[\Gamma_{\zeta_1,\zeta_2,j}\hat{\psi}_{\zeta_1}(x_j^-)\hat{\psi}_{\zeta_2}(x_j^+)+h.c.\right].\label{eqn:dH}
    \end{equation}
    
In Equation~(\ref{eqn:dH}), $j=\pm 1$ labels the right/left superconductor, and $\zeta_i$ are collective indices: $\zeta_i=\{\rho_i,\tau_i\}$. Moreover, we have introduced $x_j^{\pm}=jL/2\pm\delta_{\zeta_1,\zeta_2}\xi/2$, where $\xi=\hbar v_F/\Delta$ is the coherence length in the edges, which will be the short distance cutoff of our system, $\xi\ll L$, and $\Delta$ is the superconducting gap. This splitting makes processes with tunneling of spin-triplet CPs into or out from the same edge possible without violating the Pauli principle. It is here evident the general appeal of proximity-induced superconductivity: the symmetry of the superconducting pairing in the proximitized system can be different from the original one in the proximitizing superconductor. Finally, \mbox{the last} summation in Equation~(\ref{eqn:dH}) has been antisymmetrized for each $\zeta_1\neq\zeta_2$ term, so that $\Gamma_{\zeta_1,\zeta_2,j}=\gamma_{\zeta_1,\zeta_2,j}$ for $\zeta_1=\zeta_2$, and $\Gamma_{\zeta_1,\zeta_2,j}=\gamma_{\zeta_1,\zeta_2,j}-\gamma_{\zeta_2,\zeta_1,j}$ for $\zeta_1\neq\zeta_2$. Following this requirement, the last sum runs over only 10 terms, whose coefficients read as \cite{loss}
\vspace{-3pt}
    \begin{equation}
        \Gamma_{\zeta_1,\zeta_2,j}=\Gamma(-1)^{\delta_{\zeta_1,\{-1,-1\}}\delta_{\zeta_2,\{1,-1\}}}\left({\Tilde{f}}_T\right)^{\delta_{\rho_1\tau_1,\ \rho_2\tau_2}}(f_C)^{\delta_{\tau_1,-\tau_2}}e^{i[\frac{j}{2}k_FL(\rho_1+\rho_2)-\gamma^0_j]}.\label{eqn:gammaloss}
    \end{equation}
    
\begin{figure}[H]
	   \includegraphics[width=.95\textwidth]{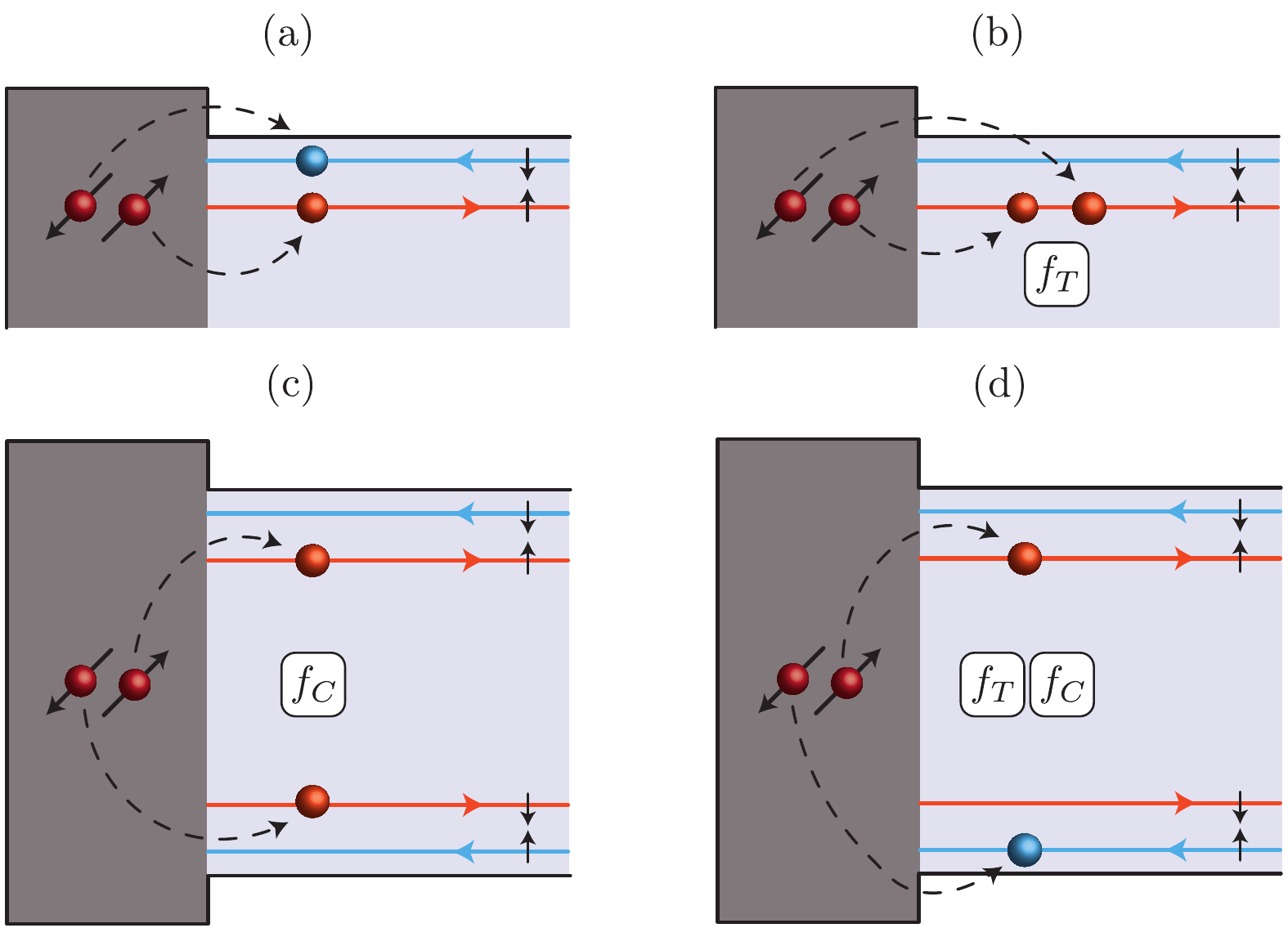}
	   \caption{Possible tunneling injections of Cooper pairs (CPs) from the superconductors to the edges of the 2DTI: either a single edge (panels (\textbf{a}) and (\textbf{b})) or both edges (panels (\textbf{c}) and (\textbf{d})) might be involved. \mbox{In the} proximitized systems, CPs can flow either in a singlet (panels (\textbf{a}) and (\textbf{c})) or in a triplet (panels (\textbf{b}) and (\textbf{d})) spin state. Triplet injections and crossed Andreev reflections affect the tunneling amplitudes with a factor of $f_T$ and $f_C$, respectively.}
	   \label{fig:tunneling}
    \end{figure}

Here, $\Gamma=\pi t^2 N_S$, where $t$ parametrizes the magnitude of tunneling across the superconductor/TI interface and $N_S$ is the normal density of states per spin in the superconductors at the Fermi energy, is the typical amplitude of the CP tunneling processes, $\gamma^0_j$ are the phases of the superconductors ($j=\pm 1$), $k_F$ is the Fermi momentum, $\Tilde{f}_T=f_T/\sqrt{1+f_T^2}$ and $f_{T/C}$ are coefficients related to the occurrence of spin flips \cite{fT} and tunneling into different edges (crossed Andreev reflection) \cite{fC}, respectively. Of particular interest for the following is $f_T$, since it allows for the propagation of triplet CPs on a single edge and hence implements processes lacking inversion symmetry. A finite $f_T$, expected due to the strong spin--orbit coupling characterizing HgTe quantum wells, is hence essential for the occurrence of the $\varphi_0$ Josephson effect in the system. Although expected as well, $f_C$ is not crucial for the following. As an only comment about such term, it is here worth mentioning that $f_C$ is expected to become significant if a perpendicular magnetic field is piercing the junction. Indeed, the presence of $f_C$ affects the periodicity of the critical supercurrent as a function of the magnetic flux \cite{njp}, since it implements a physical behavior similar to one of the so-called nano-SQUIDs \cite{nanosq}. The main classes of superconducting tunnelings are schematically shown in Figure \ref{fig:tunneling}.
\vspace{-6pt}
Having set the Hamiltonian of the system up, we can now move to the main object of our investigation: the DC Josephson current. To perform the calculation, we rely on a standard perturbative Kubo-like approach that takes into account the Hamiltonian of the edge channels as the unperturbed system and the tunneling from the superconductors up to order $\Gamma^2$, the lowest non-zero contribution (see \cite{njp} for all the details of the method). The calculation, lengthy and cumbersome but straightforward in essence, leads to the \mbox{Josephson current}
\vspace{-3pt}
	\begin{equation}
		I^{tot}=\mathcal{C}\,\text{Im}\bigg\{e^{-i\gamma_0}\mathcal{I}(f_1,f_2)-e^{i\gamma_0}\mathcal{I}(f_2,f_1)\bigg\}.\label{eqn:currentDC}
	\end{equation}
	
The function $\mathcal{I}$ has the following structure:

    \begin{align}
        \mathcal{I}(\Tilde{f}_1,\Tilde{f}_2)&=\frac{1}{{\Gamma}^2}\sum_{\zeta_1,\zeta_2,\zeta_3,\zeta_4} \Gamma_{\zeta_1,\zeta_2,1}\bar{\Gamma}_{\zeta_3,\zeta_4,-1}\int_0^{+\infty}\,ds\,\text{Im}\left[\Pi_{\zeta_1,\zeta_2,\zeta_3,\zeta_4}\left(\Tilde{L},\Tilde{f}_1,\Tilde{f}_2,\tilde{T},s\right)\right]\,,\label{eqn:italicI}
    \end{align}
and, from now on, an overbar denotes complex conjugation. Here, the integral evaluates the amplitude associated to the transport of electrons in the initial state $\zeta_3,\zeta_4$ and in the final state $\zeta_1,\zeta_2$ through the junction, hence encoding all the interesting information. More quantitatively, $\Pi_{\zeta_1,\zeta_2,\zeta_3,\zeta_4}$ arises as a combination of products of two Green functions \mbox{(see \cite{njp}} for the details) calculated over the Hamiltonian in the absence of superconductors. \mbox{For convenience,} we have rescaled all the parameters via the introduction of the \mbox{adimensional quantities}
\vspace{-4pt}
	\begin{align*}
		\Tilde{L}&=\frac{L\Delta}{\hbar v_F},\qquad \Tilde{T}=\frac{\pi k_B T}{\Delta},\qquad
		\Tilde{f_{\nu}}=\frac{f_{\nu}}{\Delta},
	\end{align*}
with $T$ being the temperature and $k_B$ being the Boltzmann constant. In Equation~(\ref{eqn:currentDC}), the natural scale is set by $\mathcal{C}\equiv(-2e\Delta\Gamma^2)/(\pi^2\hbar^3v_F^2)$, and the phase
\vspace{-3pt}
    \begin{equation}
        \gamma_0=\gamma^0_1-\gamma^0_{-1}\label{eqn:phase}
    \end{equation}
is the difference between the bare phases of the two superconductors.


\section{Results} \label{sec:results}
We now proceed to the analysis of the most striking effect related to Equation~(\ref{eqn:currentDC}), that is the anomalous Josephson current flowing in the system when the phases $\gamma^0_j$ are set to be equal. To better characterize such an effect, we first recast the current in the form
\vspace{-4pt}
    \begin{align}
		I^{tot}=\mathcal{C}(\mathcal{A}\cos\gamma_0+\mathcal{B}\sin\gamma_0)\equiv\mathcal{C}\mathcal{D}\sin(\gamma_0+\varphi_0),\label{eqn:phi01}
	\end{align}
where $\gamma_0$ has been introduced in Equation~(\ref{eqn:phase}), and we have defined
\vspace{-4pt}
	\begin{align}
        &\mathcal{A}\equiv\text{Im}\left[\mathcal{I}\left(\tilde{f}_1,\tilde{f}_2\right)-\mathcal{I}\left(\tilde{f}_2,\tilde{f}_1\right)\right], &&\mathcal{B}\equiv-\text{Re}\left[\mathcal{I}\left(\tilde{f}_1,\tilde{f}_2\right)+\mathcal{I}\left(\tilde{f}_2,\tilde{f}_1\right)\right],\notag\\
        &\mathcal{D}\equiv\sqrt{\mathcal{A}^2+\mathcal{B}^2}, &&\tan{\varphi_0}\equiv\frac{\mathcal{A}}{\mathcal{B}}.\label{eqn:phi02}
	\end{align}
	
In Equation~(\ref{eqn:phi02}), we have hence singled out the parameter $\varphi_0$, that is the one describing the anomalous Josephson effect. When $\varphi_0=0$ mod $2\pi$ there is no anomalous Josephson effect. The effect is present otherwise. For completeness, we mention that a notable value in this context is $\varphi_0=\pi$, where one has the so-called $\pi$-junction, that has less symmetry requirements with respect to the generic case. In this work, we, however, do not put our focus on such a special case.

To give a quantitative description of the $\varphi_0$ effect we predict, we now inspect the current for a vanishing phase difference between the superconductors $\gamma_0=0$. What we explicitly have is then
\vspace{-3pt}
    \begin{equation}
        I_a=I^{tot}\|_{\gamma_0=0}=\mathcal{C}\,\text{Im}\left[\mathcal{I}\left(\tilde{f}_1,\tilde{f}_2\right)-\mathcal{I}\left(\tilde{f}_2,\tilde{f}_1\right)\right]. 
    \end{equation}
    
It is clear from this relation that processes contributing to $\mathcal{I}$ symmetrically under the exchange $\tilde{f}_1\leftrightarrow \tilde{f}_2$ do not produce an anomalous current ($I_a=0$). Moreover, we find that the processes responsible for the anomalous effect are $\propto\Tilde{f}_T^2$ and independent of $f_C$ (more details and our interpretation can be found below). They hence correspond to injections as depicted in the upper-right panel of Figure \ref{fig:tunneling}. Quantitatively, we have
\vspace{-3pt}
    \begin{equation}
         I_a=\frac{\mathcal{C}}{\Gamma^2}\,\left\{\text{Im}\left[\sum_{i=1}^8\alpha_i\int_0^{+\infty}\,ds\,\text{Im}\ \Pi_i(s)\right]-\tilde{f}_1\leftrightarrow\tilde{f}_2\right\},\label{eqn:sumanomalous}
    \end{equation}
where the coefficients $\alpha_i$ are listed in the table that follows.
    
\vspace{2pt}

    \begin{center}
        \begin{tabular}{ll}
        \toprule
        Coefficient $\alpha_i$ & Corresponding $\Gamma_{\zeta_1,\zeta_2,1}\bar{\Gamma}_{\zeta_3,\zeta_4,-1}$ \\
        \midrule
        $\alpha_1=\Gamma^2\Tilde{f}_T^2 e^{-i2k_FL}$ & $=\Gamma_{\{-1,1\},\{-1,1\},1}\bar{\Gamma}_{\{-1,1\},\{-1,1\},-1}$\\
        $\alpha_2=\Gamma^2\Tilde{f}_T^2 e^{-i2k_FL}$ & $=\Gamma_{\{-1,-1\},\{-1,-1\},1}\bar{\Gamma}_{\{-1,-1\},\{-1,-1\},-1}$\\
        $\alpha_3=\Gamma^2\Tilde{f}_T^2 e^{-i2k_FL}$ & $=\Gamma_{\{-1,1\},\{-1,1\},1}\bar{\Gamma}_{\{-1,-1\},\{-1,-1\},-1}$\\
        $\alpha_4=\Gamma^2\Tilde{f}_T^2 e^{-i2k_FL}$ & $=\Gamma_{\{-1,-1\},\{-1,-1\},1}\bar{\Gamma}_{\{-1,1\},\{-1,1\},-1}$\\
        $\alpha_5=\Gamma^2\Tilde{f}_T^2 e^{i2k_FL}$ & $=\Gamma_{\{1,1\},\{1,1\},1}\bar{\Gamma}_{\{1,1\},\{1,1\},-1}$\\
        $\alpha_6=\Gamma^2\Tilde{f}_T^2 e^{i2k_FL}$ & $=\Gamma_{\{1,-1\},\{1,-1\},1}\bar{\Gamma}_{\{1,-1\},\{1,-1\},-1}$\\
        $\alpha_7=\Gamma^2\Tilde{f}_T^2 e^{i2k_FL}$ & $=\Gamma_{\{1,1\},\{1,1\},1}\bar{\Gamma}_{\{1,-1\},\{1,-1\},-1}$\\
        $\alpha_8=\Gamma^2\Tilde{f}_T^2 e^{i2k_FL}$ & $=\Gamma_{\{1,-1\},\{1,-1\},1}\bar{\Gamma}_{\{1,1\},\{1,1\},-1}$\\
        \bottomrule
        \end{tabular}
    \end{center}
    
It can be shown that
\vspace{-4pt}
    \begin{align*}
        \Pi_1(s)=\Pi_2(s) \qquad \Pi_3(s)=\Pi_4(s) \qquad \Pi_5(s)=\Pi_6(s) \qquad \Pi_7(s)=\Pi_8(s),
    \end{align*}
and that, more generally, they can all be written in a unified form. The explicit form of $\Pi_{1/3/5/7}(s)$ and other computational details are provided in Appendix \ref{sec:appendix} for clarity. All in all, the anomalous current characterizing the system reads as
\vspace{-4pt}
    \begin{align}
        I_a = \mathcal{C}\mathcal{D}\sin(\varphi_0)= 4 \mathcal{C} \Tilde{f}^2_T \sin (2 k_F L) \text{Im} \left\{ [\cosh (\lambda_2) - \cosh (\lambda_1)] \left[ \mathcal{F} \left(\Tilde{L}, \Tilde{T}\right) - \frac{2}{3} \Tilde{T}^3 \right] \right\},\label{eqn:finalanomalous}
    \end{align}
with the function $\mathcal{F}\left(\Tilde{L},\Tilde{T}\right)$ given by
\vspace{-4pt}
    \begin{equation*}
        \mathcal{F}\left(\Tilde{L},\Tilde{T}\right)=\frac{4}{3} \Tilde{T}^3 \frac{1 - 3 e^{2\left(\Tilde{L} - i\right) \Tilde{T}}}{\left[1 - e^{2 \left(\Tilde{L} - i\right) \Tilde{T}}\right]^3}\qquad\text{and}\qquad\lambda_{\nu} = 2 \Tilde{f}_{\nu} \left(1 + i \Tilde{L}\right),\,\nu=1,2.
    \end{equation*}


\section{Discussion} \label{sec:discussion}
\begin{figure}[H]
		\includegraphics[width=\textwidth]{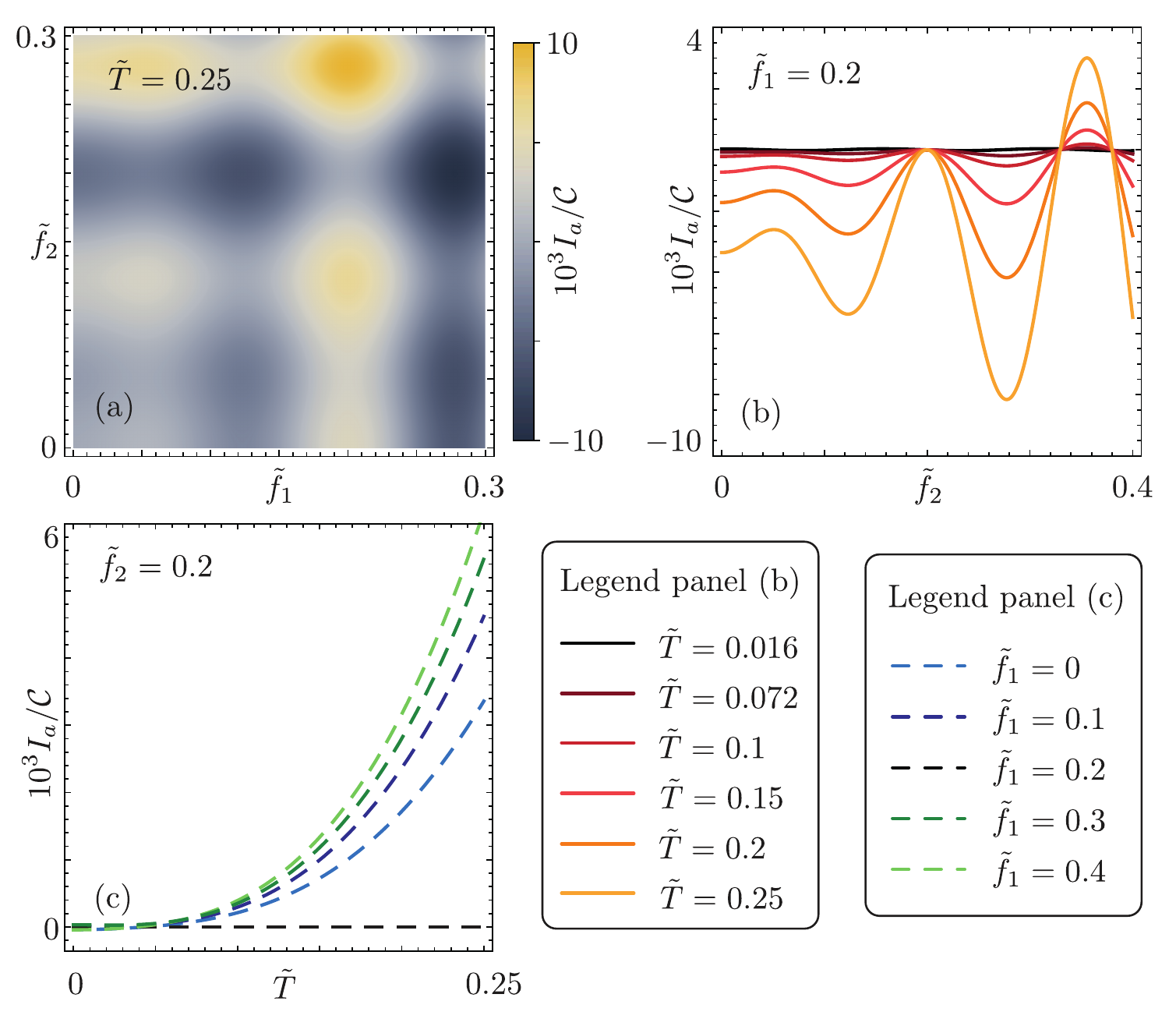}
		\caption{Anomalous current $I_a=I^{tot}\|_{\gamma_0=0}$ flowing in the system. (\textbf{a}) $I_a/\mathcal{C}$ as a function of $\Tilde{f}_1,\,\Tilde{f}_2$, with $f_C=0.3$, $\Tilde{f}_T=0.4$, $k_FL=23/4\pi$, $\Tilde{L}=20$ and $\Tilde{T}=0.25$. (\textbf{b}) $I_a/\mathcal{C}$ as a function of $\Tilde{f}_2$, for different values of $\Tilde{T}$ (see the plot legend) and $\Tilde{f}_1=0.2$. The other parameters remain unchanged from (\textbf{a}). (\textbf{c}) $I_a/\mathcal{C}$ as a function of $\Tilde{T}$, for different values of $\Tilde{f}_1,\,\Tilde{f}_2$ (see the plot legend). The other parameters remain unchanged from (\textbf{a}).}
		\label{fig:anomalous}
	\end{figure}

The anomalous current is shown in Figure \ref{fig:anomalous}, as a function of temperature and tunneling amplitudes.

First of all, we note that the anomalous current does not have a well-defined sign. Indeed, as expected, it is odd under the exchange $\tilde{f}_1\leftrightarrow \tilde{f}_2$, which essentially realizes the inversion of the structure. Accordingly, we find $I_a=0$ for $\tilde{f}_1=\tilde{f}_2$, and in that case, Equation~(\ref{eqn:phi01}) recovers the standard Josephson effect. This behavior could be expected since for $\tilde{f}_1=\tilde{f}_2$ time-reversal symmetry is preserved. An oscillatory behavior is also present as a function of $\tilde{f}_1$ and $\tilde{f}_2$. This fact is not surprising since oscillations in the amplitude of the DC and AC (non-anomalous) Josephson currents are also present for $\tilde{f}_1=\tilde{f}_2$ \cite{njp} (Figure \ref{fig:anomalous}a). As a second interesting fact, thanks to such oscillations, $I_a$ increases rapidly as soon as an asymmetry between the couplings, although small, is introduced. This is visible in \mbox{Figure \ref{fig:anomalous}b,} where we set $\tilde{f}_1=0.2$: the anomalous current assumes a finite value as $\tilde{f}_2$ deviates, even slightly, from 0.2. Finally, it is here worth noticing a peculiar effect: the anomalous current increases as the temperature is increased, keeping, however, $T$ smaller than the induced gap (Figure \ref{fig:anomalous}c). This effect can be understood with a simple ``perturbative'' argument: the scattering between the edges, reported in Equation~(\ref{eqn:fs}), mixes states with the same kinetic energy according to Equation~(\ref{eqn:kin}). Hence, in a Fermi's golden rule approach, its effects are strongly suppressed at low temperature due to the Fermi functions. Such a temperature dependence is hence a clear cut signature of the constriction. Lastly, we point out that, going beyond the low-transparency assumption, and hence including higher order contributions of $\delta\hat{H}_E$ in the calculation of the current, would result in a superconducting diode effect \cite{klinovaja}.

In order to capture the essential physics underlying the temperature behavior, some expansions of Equation~(\ref{eqn:finalanomalous}) are particularly helpful. First of all, the anomalous current has a zero-temperature contribution, although hardly visible in Figure \ref{fig:anomalous}c:

    \begin{equation}
        \left.{\frac{I_a}{\mathcal{C} \Tilde{f}^2_T \sin (2 k_F L)}}\right\|_{\Tilde{T}=0}=\frac{4}{3} \text{Im} \left\{
   \frac{\cosh (\lambda_2) - \cosh (\lambda_1)}{\left(\Tilde{L} - i\right)^3} \right\}.\label{eqn:zeroT}
    \end{equation}
    
Restoring finite temperature, and in the long junction limit $\Tilde{T}\Tilde{L}\gg1$, we obtain
\vspace{-4pt}
{\small{
    \begin{align}
        \frac{I_a}{\mathcal{C} \Tilde{f}^2_T \sin (2 k_F L)} \overset{\Tilde{T}\Tilde{L}\gg1}{\simeq}&16\left[\cos\left(2\Tilde{f}_2\Tilde{L}\right)\cosh\left(2\Tilde{f}_2\right)-\cos\left(2\Tilde{f}_1\Tilde{L}\right)\cosh\left(2\Tilde{f}_1\right)\right]\Tilde{T}^3e^{-4\Tilde{L}\Tilde{T}}\sin\left(4\Tilde{T}\right)\notag\\
        -&\frac{8}{3}\left[\sin\left(2\Tilde{f}_2\Tilde{L}\right)\sinh\left(2\Tilde{f}_2\right)-\sin\left(2\Tilde{f}_1\Tilde{L}\right)\sinh\left(2\Tilde{f}_1\right)\right]\Tilde{T}^3.\label{eqn:TL}
    \end{align}}}
    
On the other hand, without assumptions on $\Tilde{L}$ but for small $\Tilde{T}$ and $\Tilde{f_{\nu}}$, we can further expand Equation~(\ref{eqn:finalanomalous}) as
\vspace{-4pt}
    \begin{align}
        \frac{I_a}{\mathcal{C} \Tilde{f}^2_T \sin (2 k_F L)} &\overset{\Tilde{T},\Tilde{f}_{\nu}\ll1}{\simeq}\frac{8}{3} \left(\Tilde{f}_1^2 - \Tilde{f}_2^2\right) \left[\frac{1}{1 + \Tilde{L}^2} + 2 \Tilde{T}^2 \right],\label{eqn:finiteT} 
    \end{align}
that is the temperature scaling visible in Figure \ref{fig:anomalous}c.

We are now in the position of interpreting the temperature dependence of the anomalous current. At zero temperature (see Equation~(\ref{eqn:zeroT})), for small $\tilde{f}_{1/2}$, the inter-edge tunneling events are suppressed by phase space arguments. The coupling between the edges, \linebreak for $\tilde{f}_1\neq\tilde{f}_2$, hence only splits the energies of electrons with different chiralities. In terms of the spin degree of freedom, it is thus analogous to the application of two magnetic fields in the $z$-direction opposite to each other on the two edges. It hence generates a $\varphi_0$ effect analogous of the one reported in \cite{dolcini}. However, thanks to the fact that the effect is now opposite for the two edges, the anomalous Josephson current does not vanish when both edges are taken into account. The energy splitting just mentioned is independent of the length $\tilde{L}$ of the system and, consistently, the associated current decays as the distance between the superconductors is increased (see Equation~(\ref{eqn:TL})), as usual for Josephson currents. When the temperature is finite (see Equation~(\ref{eqn:finiteT})), a different mechanism for the $\varphi_0$ effect is added to the one just discussed: The inter-edge tunneling events. Just as the zero temperature contribution, the finite temperature one scales as $\tilde{f}_{1/2}^2$. It thus involves two inter-edge tunneling events. The probability of each tunneling event scales as $\tilde{L}$, so that the finite temperature term scales, with respect to the zero temperature one, with a factor $\tilde{L}^2$ more. Moreover, each tunneling corresponds to a product $\mathsf{f}(1-\mathsf{f})$, with $\mathsf{f}$ the Fermi function. Under integration, such a product contributes with a factor $\tilde{T}$, and since we are in the presence of a double tunneling, the leading order becomes $\tilde{T}^2$. To summarize, the probability of the events contributing to the temperature activated anomalous current increases with both the junction length and temperature. Our qualitative interpretation is hence able to capture the behavior of the $\varphi_0$ effect at finite temperature, its non-zero value at zero temperature, and the fact that the temperature activated processes have a better scaling with the length of the junction.

\section{Conclusions}\label{sec:conc}
In this article, we have conceived an experimentally relevant system presenting a $\varphi_0$ Josephson effect in the absence of applied magnetic fields. Such a system is a JJ with a reconstructed topological constriction as a link. We have then analytically assessed the anomalous Josephson current which, surprisingly at first, increases as the temperature is increased. We have analyzed in detail such an increase by means of a perturbative expansion and we have qualitatively interpreted that on the basis of the thermal activation of the tunneling processes in the constriction. Our results open the way to the design of phase batteries in the system we analyze, remarkably in the absence of external magnetic fields, since the required time-reversal symmetry breaking is provided by the edge reconstruction. In addition, the fact that the main building block of our system is a 2DTI makes it possible to envision the direct integration of the $\varphi_0$ junction we considered with other functional nanostructures built on the same TI. Lastly, when the limit of weak tunneling between the system and the superconductors is relaxed, we expect our setup to show a superconducting diode effect, based on symmetry arguments. 

\vspace{6pt} 



\authorcontributions{Conceptualization, L.V. and N.T.Z.; investigation, L.V., F.C. and N.T.Z.; validation, L.V., F.C., G.P., M.S. and N.T.Z.; writing---original draft, N.T.Z.; writing---review and editing, L.V., F.C., G.P., M.S. and N.T.Z. All authors have read and agreed to the published version of the manuscript.}

\funding{Funding through the NextGenerationEu Curiosity Driven Project ``Understanding even-odd criticality'' is acknowledged.}

\institutionalreview{Not applicable.}

\informedconsent{Not applicable.}

\dataavailability{Not applicable.} 

\acknowledgments{L.V. acknowledges E. Bocquillon for helpful discussions at the early stage of \mbox{the project.}}

\conflictsofinterest{The authors declare no conflicts of interest.} 


\abbreviations{Abbreviations}{
The following abbreviations are used in this manuscript:\\

\noindent 
\begin{tabular}{@{}ll}
JJ & Josephson junction\\
2D & two-dimensional\\
TI & topological insulator\\
SQUID & superconducting quantum interference device\\
CP & Cooper pair
\end{tabular}
}

\appendixtitles{yes} 
\appendixstart
\appendix
\section[\appendixname~\thesection]{Main Steps to obtain Equation~(\ref{eqn:finalanomalous})} \label{sec:appendix}
In this Appendix, we provide some more details about the connection between \mbox{Equations~(\ref{eqn:sumanomalous}) and (\ref{eqn:finalanomalous}).} Let us start from Equation~(\ref{eqn:sumanomalous}), and let us define the function
\vspace{-4pt}
    \begin{equation*}
        K (s) = \frac{1}{\sinh{\left[\Tilde{T}\left(\tilde{L}+s-i\right)\right]}},\label{eq:defK}
    \end{equation*}
with $s\in\mathbb{R}$. Moreover, let us recall the definition
\vspace{-4pt}
    \begin{equation*}
        \lambda_{\nu} = 2 \Tilde{f}_{\nu} \left(1 + i \Tilde{L}\right),\,\nu=1,2
    \end{equation*}
 already given in the main text. From calculations that we skip here, but can be inferred by following \cite{njp}, the functions $\Pi_{1/3/5/7}(s)$ turn out to be given by
\vspace{-4pt}
{\small{
    \begin{align}
        \!\!\!\!\!\!&\Pi_1(s)=\frac{\Tilde{T}^2}{2}\left\{[\cosh(\lambda_2)+ 1]K^2(s)-\left[\cosh(\lambda_2)+\cos\left(2\Tilde{f}_2\right)\right]K(s+1)K(s-1)\right\},\\
        \!\!\!\!\!\!&\Pi_3(s)=\frac{\Tilde{T}^2}{2}\left\{[\cosh(\lambda_2)-1]K^2(s)-\left[\cosh(\lambda_2)-\cos\left(2\Tilde{f}_2\right)\right]K(s+1)K(s-1)\right\},\\
        \!\!\!\!\!\!&\Pi_5(s)\!=\!\frac{\Tilde{T}^2}{2}\left\{[\cosh(\lambda_1)+ 1]\bar{K}^2(-s)\!-\!\left[\cosh(\lambda_1)+\cos\left(2\Tilde{f}_1\right)\right]\bar{K}(-s-1)\bar{K}(-s+1)\right\},\\
        \!\!\!\!\!\!&\Pi_7(s)\!=\!\frac{\Tilde{T}^2}{2}\left\{[\cosh(\lambda_1)- 1]{\bar{K}}^2(-s)-\left[\cosh(\lambda_1)\!-\!\cos\left(2\Tilde{f}_1\right)\right]\bar{K}(-s-1)\bar{K}(-s+1)\right\},
    \end{align}
    }}
where the overbar denotes the complex conjugation. By noticing the symmetry relation

    \begin{equation*}
        \bar{K}(-s)=\frac{1}{\sinh\left[\Tilde{T}\left(\Tilde{L}-s+i\right)\right]}\equiv K(-s+2i),
    \end{equation*}
we can rewrite the functions $\Pi_5(s)$ and $\Pi_7(s)$ as
\vspace{-4pt}
    \begin{align}
        \Pi_5(s)=&\frac{\Tilde{T}^2}{2}\bigg\{[\cosh(\lambda_1)+1]K^2(-s+2i)-\left[\cosh(\lambda_1)+\cos\left(2\Tilde{f}_1\right)\right]\notag\\
        &K(-s+2i-1)K(-s+2i+1)\bigg\},\\
        \Pi_7(s)=&\frac{\Tilde{T}^2}{2}\bigg\{[\cosh(\lambda_1)-1]K^2(-s+2i)-\left[\cosh(\lambda_1)-\cos\left(2\Tilde{f}_1\right)\right]\notag\\
        &K(-s+2i-1)K(-s+2i+1)\bigg\}. 
    \end{align}
Finally, in a unified form, we obtain
\vspace{-4pt}
    \begin{align}
        \Pi_1(s) \equiv \Pi^{(2)}_+ (s); \ \Pi_3(s) \equiv \Pi^{(2)}_- (s); \ \Pi_5(s) \equiv \Pi^{(1)}_+ (- s + 2 i); \ \Pi_7(s) \equiv \Pi_-^{(1)} (- s + 2 i)
    \end{align}
where
\vspace{-3pt}
    \begin{equation}
        \Pi_{\pm}^{(\nu)}(s)=\frac{\Tilde{T}^2}{2}\left\{[\cosh(\lambda_{\nu})\pm1]K^2(s)-\left[\cosh(\lambda_{\nu})\pm\cos\left(2\Tilde{f}_{\nu}\right)\right]K(s+1)K(s-1)\right\}.\label{eqn:pi+-}
    \end{equation}
    
Since $\xi$ is the short distance cut-off in our model, and since we are interested in the long junction regime, we can expand the functions $K$ at the lowest order in $\xi/L$. In this case, what we obtain is
\vspace{-4pt}
\begin{equation}
       K(s+1)K(s-1)\simeq\Tilde{T}^2K^4(s)+K^2(s),
    \end{equation}
and hence Equation~(\ref{eqn:pi+-}) becomes
\vspace{-4pt}
\begin{equation}
        \Pi_{\pm}^{(\nu)}(s)=\frac{\Tilde{T}^2}{2}\left\{\left[\mp\cos\left(2\Tilde{f}_{\nu}\right)\pm1\right]K^2(s)-\left[\cosh(\lambda_{\nu})\pm\cos\left(2\Tilde{f}_{\nu}\right)\right]\Tilde{T}^2K^4(s)\right\}.
    \end{equation}
    
Plugging this result into Equation~(\ref{eqn:sumanomalous}) and extracting the imaginary parts, the terms $\propto K^2(s)$ cancel out exactly and we are left with the following expression for the anomalous current, which represents the central result of our article:
\vspace{-4pt}
    \begin{align}
        \frac{I_a}{\mathcal{C}}=&-2\tilde{f}_T^2\sin{(2k_FL)}\left[\cosh{(2\tilde{f}_1)}\cos{(2\tilde{f}_1\tilde{L})}\left(I_1+I_2\right)\right.\notag\\
        &\left.-\sinh{(2\tilde{f}_1)}\sin{(2\tilde{f}_1\tilde{L})}\left(I_4-I_3\right)+\tilde{f}_1\leftrightarrow\tilde{f}_2\right],\label{eqn:simplifiedanomalous}
    \end{align}

\noindent where
\vspace{-4pt}
    \begin{align*}
        &I_1=\tilde{T}^4\,\text{Im}\left[\int_0^{+\infty}\,ds\,K^4(s)\right],\quad
        I_2=\tilde{T}^4\,\text{Im}\left[\int_0^{+\infty}\,ds\,\bar{K}^4(-s)\right],\\
        &I_3=\tilde{T}^4\,\text{Re}\left[\int_0^{+\infty}\,ds\,K^4(s)\right],\quad I_4=\tilde{T}^4\,\text{Re}\left[\int_0^{+\infty}\,ds\,\bar{K}^4(-s)\right].
    \end{align*}

The building block here is given by
\vspace{-4pt}
{\small{
    \begin{equation*}
        \tilde{T}^4\int_0^{+\infty}\,ds\,K^4(s)=\frac{4}{3} \tilde{T}^3 \frac{1 - 3 e^{2 \left(\Tilde{L} - i\right) \Tilde{T}}}{\left[1 - e^{2 \left(\Tilde{L} - i\right) \Tilde{T}}\right]^3}\equiv\mathcal{F} \left(\Tilde{L},\Tilde{T}\right),\qquad \tilde{T}^4\int_0^{+\infty}\,ds\,\bar{K}^4(-s)=\mathcal{F} \left(-\Tilde{L},\Tilde{T}\right),
    \end{equation*}}}
and we notice that
\vspace{-4pt}
    \begin{equation*}
        I_1 = I_2 \qquad I_3 + I_4 = \frac{4}{3} \Tilde{T}^3.
    \end{equation*}
    
By means of some more simple algebraic manipulations, one can now finally reach the final expression give in Equation~(\ref{eqn:finalanomalous}).

\begin{adjustwidth}{-\extralength}{0cm}

\reftitle{References}

\PublishersNote{}
\end{adjustwidth}

\begin{thebibliography}{999}

\bibitem{tinkham}
    Tinkham, M. \textit{Introduction to Superconductivity}; McGraw-Hill: New York, NY, USA, {1996}.

\bibitem{barone}
    Barone, A.; Paternò, G. \textit{Physics and Applications of the Josephson Effect}; John Wiley \& Sons, Ltd.: Hoboken, NJ, USA, {1982}.

\bibitem{nanosq}
    Granata, C.; Vettoliere, A. Nano Superconducting Quantum Interference device: A powerful tool for nanoscale investigations. {\em Phys. Rep.} {\bf 2016}, {\em 614}, 1--69.

\bibitem{supq}
    Clarke, J.; Wilhelm, F.K. Superconducting quantum bits. {\em Nature} {\bf 2008}, {\em 453}, 1031--1042.

\bibitem{kitaev}
    Kitaev, A.Y. Unpaired Majorana fermions in quantum wires. {\em Phys. Usp.} {\bf 2001}, {\em 44}, 131.

\bibitem{maj2}
    Albrecht, S.M.; Higginbotham, A.P.; Madsen, M.; Kuemmeth, F.; Jespersen, T.S.; Nyg\aa rd, J.; Krogstrup, P.; Marcus, C.M. Exponential protection of zero modes in Majorana islands. {\em Nature} {\bf 2016}, {\em 531}, 206--209.

\bibitem{prb}
    Traverso Ziani, N.; Fleckenstein, C.; Vigliotti, L.; Trauzettel, B.; Sassetti, M. From fractional solitons to Majorana fermions in a paradigmatic model of topological superconductivity. {\em Phys. Rev. B} {\bf 2020}, {\em 101}, 195303. 

\bibitem{para}
    Clarke, D.J., Alicea, J., Shtengel, K. Exotic non-Abelian anyons from conventional fractional quantum Hall states. {\em Nat. Commun.} {\bf 2013}, {\em 4}, 1348.

\bibitem{para2}
    Fleckenstein, C.; Traverso Ziani, N.; Trauzettel, B. $\mathbb{Z}_4$ parafermions in Weakly Interacting Superconducting Constrictions at the Helical Edge of Quantum Spin Hall Insulators. {\em Phys. Rev. Lett.} {\bf 2019}, {\em 122}, 066801.

\bibitem{sst}
    Linder, J.; Robinson, J.W.A. Superconducting spintronics. {\em Nat. Phys.} {\bf 2015}, {\em 11}, 307--315.

\bibitem{calo}
    Fornieri, A.; Giazotto, F. Towards phase-coherent caloritronics in superconducting circuits. {\em Nat. Nanotechnol.} {\bf 2017}, {\em 12}, 944--952.

\bibitem{symmetry0}
    Golubov, A.A.; Kupriyanov, M.Y.; Il’ichev, E. The current-phase relation in Josephson junctions. {\em Rev. Mod. Phys.} {\bf 2004}, {\em 76}, 411.

\bibitem{symmetry}
    Rasmussen, A.; Danon, J.; Suominen, H.; Nichele, F.; Kjaergaard, M.; Flensberg, K. Effects of spin-orbit coupling and spatial symmetries on the Josephson current in SNS junctions. {\em Phys. Rev. B} {\bf 2016}, {\em 93}, 155406.

\bibitem{symmetry2}
    Liu, J.-F.; Chan, K.S. Relation between symmetry breaking and the anomalous Josephson effect. {\em Phys. Rev. B} {\bf 2010}, {\em 82}, 125305.

\bibitem{shukrinov}
    Shukrinov, Y.M. Anomalous Josephson effect. {\em Phys. Usp.} {\bf 2022}, {\em 65}, 317.

\bibitem{phizero}
    Buzdin, A. Direct coupling between magnetism and superconducting current in the Josephson $\phi_0$ junction, {\em Phys. Rev. Lett.} {\bf 2008} {\em 101}, 107005.

\bibitem{phizero1}
    Baumgartner, C.; Fuchs, L.; Costa, A.; Picó-Cortés, J.; Reinhardt, S.; Gronin, S.; Gardner, G.C.; Lindemann, T.; Manfra, M.J.; Faria Junior, P.E.; Kochan, D.; Fabian, J.; Paradiso, N.; Strunk, C. Effect of Rashba and Dresselhaus spin–orbit coupling on supercurrent rectification and magnetochiral anisotropy of ballistic Josephson junctions. {\em J. Phys. Condens. Matter} {\bf 2022}, {\em 34}, 154005.

\bibitem{phizero2}
    Goldobin, E.; Koelle, D.; Kleiner, R.; Mints, R.G. Josephson Junction with a Magnetic-Field Tunable Ground State. {\em Phys. Rev. Lett.} {\bf 2011}, {\em 107}, 227001.

\bibitem{phizero3}
    Silaev, M.A.; Tokatly, I.V.; Bergeret, F.S. Anomalous current in diffusive ferromagnetic Josephson junctions. {\em Phys. Rev. B} {\bf 2017}, \mbox{{\em 95}, 184508.}

\bibitem{phizero4}
    Szombati, D.B.; Nadj-Perge, S.; Car, D.; Plissard, S.R.; Bakkers, E.P.A.M.; Kouwenhoven, L.P. Josephson $\varphi_0$-junction in nanowire quantum dots. {\em Nat. Phys.} {\bf 2016}, {\em 12}, 568–572.

\bibitem{phizero5}
    Assouline, A.; Feuillet-Palma, C.; Bergeal, N.; Zhang, T.; Mottaghizadeh, A.; Zimmers, A.; Lhuillier, E.; Eddrie, M.; Atkinson, P.; Aprili, M.; Aubin, H. Spin-Orbit induced phase-shift in $Bi_2Se_3$ Josephson junctions. {\em Nat. Commun.} {\bf 2019}, {\em 10}, 126.

\bibitem{phizero6}
    Mayer, W.; Dartiailh, M.C.; Yuan, J.; Wickramasinghe, K.S.; Rossi, E.; Shabani, J. Gate controlled anomalous phase shift in Al/InAs Josephson junctions. {\em Nat. Commun.} {\bf 2020}, {\em 11}, 212.

\bibitem{alidoust1}
    Alidoust, M.; Hamzehpour, H. Spontaneous supercurrent and $\varphi_0$ phase shift parallel to magnetized topological insulator interfaces. {\em Phys. Rev. B} {\bf 2017}, {\em 96}, 165422.

\bibitem{alidoust2}
    Alidoust, M. Critical supercurrent and $\varphi_0$ state for probing a persistent spin helix. {\em Phys. Rev. B} {\bf 2020}, {\em 101}, 155123.
    
\bibitem{diode0}
    Zhang, Y.; Gu, Y.; Li, P.; Hu, J.; Jiang, K. General Theory of Josephson Diodes. {\em Phys. Rev. X} {\bf 2022}, {\em 12}, 041013.

\bibitem{diode}
    Hu, J.; Wu, C.; Dai, X. Proposed Design of a Josephson Diode. {\em Phys. Rev. Lett.} {\bf 2007}, {\em 99}, 067004.

\bibitem{diode1}
    Fu, P.-H.; Xu, Y.; Lee, C.H.; Ang, Y.S.; Liu, J.-F. Gate-Tunable High-Efficiency Topological Josephson Diode. {\em arXiv} {\bf 2022}, arXiv:2212.01980.

\bibitem{alidoust3}
    Halterman, K.; Alidoust, M.; Smith, R.; Starr, S. Supercurrent diode effect, spin torques, and robust zero-energy peak in planar half-metallic trilayers. {\em Phys. Rev. B} {\bf 2022}, {\em 105}, 104508.

\bibitem{alidoust4}
    Alidoust, M.; Willatzen, M.; Jauho, A.-P. Fraunhofer response and supercurrent spin switching in black phosphorus with strain and disorder. {\em Phys. Rev. B} {\bf 2018}, {\em 98}, 184505.

\bibitem{phizeroexp}
    Sickinger, H.; Lipman, A.; Weides, M.; Mints, R.G.; Kohlstedt, H.; Koelle, D.; Kleiner, R.; Goldobin, E. Experimental Evidence of a $\varphi$ Josephson Junction. {\em Phys. Rev. Lett.} {\bf 2012}, {\em 109}, 107002.

\bibitem{battery}
    Strambini, E.; Iorio, A.; Durante, O.; Citro, R.; Sanz-Fern{\'a}ndez, C.; Guarcello, C.; Tokatly, I.V.; Braggio, A.; Rocci, M.; Ligato, N.; Zannier, V.; Sorba, L.; Bergeret, F.S.; Giazotto, F. A Josephson phase battery. {\em Nat. Nanotechnol.} {\bf 2020}, {\em 15}, 656--660.

\bibitem{battery2}
    Pal, S.; Benjamin, C. Quantized Josephson phase battery. {\em EPL} {\bf 2019}, {\em 126}, 57002.

\bibitem{circuits}
     Virtanen, P.; Bergeret, F.S.; Strambini, E.; Giazotto, F.; Braggio, A. Majorana bound states in hybrid two-dimensional Josephson junctions with ferromagnetic insulators. {\em Phys. Rev. B} {\bf 2018}, {\em 98}, 020501.

\bibitem{memory}
    Guarcello, C.; Bergeret, F.S. Cryogenic Memory Element Based on an Anomalous Josephson Junction. {\em Phys. Rev. Appl.} {\bf 2020}, \mbox{{\em 13}, 034012.}

\bibitem{diodeexp0}
    Ando, F.; Miyasaka, Y.; Li, T.; Ishizuka, J.; Arakawa, T.; Shiota, Y.; Moriyama, T.; Yanase, Y.; Ono, T. Observation of superconducting diode effect. {\em Nature} {\bf 2020}, {\em 584}, 373.

\bibitem{diodeexp1}
    Pal, B.; Chakraborty, A.; Sivakumar, P.K.; Davydova, M.; Gopi, A.K.; Pandeya, A.K.; Krieger, J.A.; Zhang, Y.; Date, M.; Ju, S.; Yuan, N.; Schr\"oter, N.B.M.; Fu, L.; Parkin, S.S.P. Josephson diode effect from Cooper pair momentum in a topological semimetal. {\em Nature Phys.} {\bf 2022}, {\em 18}, 1228--1233.

\bibitem{diodeexp2}
   Lin, J.-X.; Siriviboon, P.; Scammell, H.D.; Liu, S.; Rhodes, D.; Watanabe, K.; Taniguchi, T.; Hone, J.; Scheurer, M.S.; Li, J.I.A. Zero-field superconducting diode effect in small-twist-angle trilayer graphene. {\em Nat. Phys.} {\bf 2022}, {\em 18}, 1221--1227.

\bibitem{diode2}
    Jiang, K.; Hu, J. Superconducting diode effects. {\em Nat. Phys.} {\bf 2022}, {\em 18}, 1145--1146.

\bibitem{referee}
    Cefalas, A.C.; Kollia, Z.; Spyropoulos-Antonakakis, N.; Gavriil, V.; Christofilos, D.; Kourouklis, G.; Semashko, V.V.; Pavlov, V.; Sarantopoulou, E. Surface profile gradient in amorphous $Ta_2O_5$ semi conductive layers regulates nanoscale electric current stability. {\em Appl. Surf. Sci.} {\bf 2017}, {\em 396}, 1000--1019.

\bibitem{hou}
    Hou, Y.; Nichele, F.; Chi, H.; Lodesani, A.; Wu, Y.; Ritter, M.F.; Haxell, D.Z.; Davydova, M.; Ilić, S.; Bergeret, F.S.; Kamra, A.; Fu, L.; Lee, P.A.; Moodera, J.S. Ubiquitous superconducting diode effect in superconductor thin films. {\em arXiv} {\bf 2022}, arXiv:2205.09276.

\bibitem{seoane}
    Souto, R.S.; Leijnse, M.; Schrade, C. Josephson Diode Effect in Supercurrent Interferometers. {\em Phys. Rev. Lett.} {\bf 2022}, {\em 129}, 267702.

\bibitem{inversion}
    Chen, C.-Z.; He, J.J.; Ali, M.N.; Lee, G.-H.; Fong, K.C.; Law, K.T. Asymmetric Josephson effect in inversion symmetry breaking topological materials. {\em Phys. Rev. B} {\bf 2018}, {\em 98}, 075430.

\bibitem{noncentrosymm}
    Wakatsuki, R.; Saito, Y.; Hoshino, S.; Itahashi, Y.M.; Ideue, T.; Ezawa, M.; Iwasa, Y.; Nagaosa, N. Nonreciprocal charge transport in noncentrosymmetric superconductors, {\em Sci. Adv.} {\bf 2017}, {\em 3}, e1602390.

\bibitem{filippozzi}
    Wu H.; Wang, Y.; Xu, Y.; Sivakumar P.K.; Pasco, C.; Filippozzi, U.; Parkin, S.S.P.; Zeng, Y.J.; McQueen, T.; Ali, M.N. The field-free Josephson diode in a van der Waals heterostructure. {\em Nature} {\bf 2022}, {\em 604}, 653.

\bibitem{nonmag1}
    Alvarado, M.; Burset, P.; Yeyati A.L. Intrinsic non-magnetic $\phi_0$ Josephson junctions in twisted bilayer graphene. {\em arXiv} {\bf 2023}, arXiv:2303.07738.

\bibitem{nonmag2}
    Kokkeler, T.H.; Golubov, A.A.; Bergeret, F.S. Field-free anomalous junction and superconducting diode effect in spin-split superconductor/topological insulator junctions. {\em Phys. Rev. B} {\bf 2022}, {\em 106}, 214504.
    
\bibitem{qi}Qi, X.-L.; Zhang S.-C. Topological insulators and superconductors
{\em Rev. Mod. Phys.} {\bf 2011}, {\em 83}.

\bibitem{deacon}
    Deacon, R.S.; Wiedenmann, J.; Bocquillon, E.; Domínguez, F.; Klapwijk, T.M.; Leubner, P.; Brüne, C.; Hankiewicz, E.M.; Tarucha, S.; Ishibashi, K.; et al. Josephson Radiation from Gapless Andreev Bound States in HgTe-Based Topological Junctions. {\em Phys. Rev. X} {\bf 2017}, {\em 7}, 021011.

\bibitem{nanomaterials}
    Vigliotti, L.; Calzona, A.; Traverso Ziani, N.; Bergeret, F.S.; Sassetti, M.; Trauzettel, B. Effects of the Spatial Extension of the Edge Channels on the Interference Pattern of a Helical Josephson Junction. {\em Nanomaterials} {\bf 2023}, {\em 13}, 569.

\bibitem{bhz}
    Bernevig, B.A.; Hughes, T.L.; Zhang, S.-C. Quantum Spin Hall Effect and Topological Phase Transition in HgTe Quantum Wells. {\em Science} {\bf 2006}, {\em 314}, 1757.

\bibitem{tiexp}
    K\"onig, M.; Wiedmann, S.; Br\"une, C.; Roth, A.; Buhmann, H.; Molenkamp, L.W.; Qi, X.-L.; Zhang, S.-C. Quantum Spin Hall Insulator State in HgTe Quantum Wells. {\em Science} {\bf 2007}, {\em 318}, 766.

\bibitem{tinl}
    Roth, A.; Br\"une, C.; Buhmann, H.; Molenkamp, L.W.; Maciejko, J.; Qi, X.-L.; Zhang, S.-C. Nonlocal Transport in the Quantum Spin Hall State. {\em Science} {\bf 2009}, {\em 325}, 294--297.

\bibitem{ti}
    Hasan, M.Z.; Kane, C.L. Colloquium: Topological insulators. {\em Rev. Mod. Phys.} {\bf 2010} {\em 82}, 3045--3067.

\bibitem{nat1}
    Strunz, J.; Wiedenmann, J.; Fleckenstein, C.; Lunczer, L.; Beugeling, W.; M\"uller, V. L.; Shekhar, P.; Traverso Ziani, N.; Shamim, S.; Kleinlein, J.; Buhmann, H.; Trauzettel, B.; Molenkamp, L. W. Interacting topological edge channels. {\em Nat. Phys.} {\bf 2020}, {\em 16}, 83–88.

\bibitem{twotunn1}
    Li, J.; Pan, W.; Bernevig, B.A.; Lutchyn, R.M. Detection of Majorana Kramers Pairs Using a Quantum Point Contact. {\em Phys. Rev. Lett.} {\bf 2016}, {\em 117}, 046804.

\bibitem{twotunn2}
    Fleckenstein, C.; Traverso Ziani, N.; Calzona, A.; Sassetti, M.; Trauzettel, B. Formation and detection of Majorana modes in quantum spin Hall trenches. {\em Phys. Rev. B} {\bf 2021}, {\em 103}, 125303.

\bibitem{dolcettocimento}
    Dolcetto, G.; Sassetti, M.; Schmidt, T.L. Edge Physics in two-dimensional topological insulators. {\em Riv. Del Nuovo C.} {\bf 2016}, {\em 39}, 113.

\bibitem{reconstruction}
    Wang, J.; Meir, Y.; Gefen, Y. Spontaneous Breakdown of Topological Protection in Two Dimensions. {\em Phys. Rev. Lett.} {\bf 2017}, \mbox{{\em 118}, 046801.}

\bibitem{dolcini}
    Dolcini, F.; Houzet, M.; Meyer, J.S. Topological Josephson $\phi_0$ junctions. {\em Phys. Rev. B} {\bf 2015}, {\em 92}, 035428.

\bibitem{spinsass}
    Dolcetto, G.; Cavaliere, F.; Ferraro, D.; Sassetti, M. Generating and controlling spin-polarized currents induced by a quantum spin Hall antidot. {\em Phys. Rev. B} {\bf 2013}, {\em 87}, 085425.

\bibitem{biggio}
    Dolcetto, G.; Traverso Ziani, N.; Biggio, M.; Cavaliere, F.; Sassetti, M. Coulomb blockade microscopy of spin-density oscillations and fractional charge in quantum spin Hall dots. {\em Phys. Rev. B} {\bf 2013}, {\em 87}, 235423.

\bibitem{loss}
    Haidekker Galambos, T.; Hoffman, S.; Recher, P.; Klinovaja, J.; Loss, D. Superconducting Quantum Interference in Edge State Josephson Junctions. {\em Phys. Rev. Lett.} {\bf 2020}, {\em 125}, 157701.

\bibitem{njp}
    Vigliotti, L.; Calzona, A.; Trauzettel, B.; Sassetti, M.; Traverso Ziani, N. Anomalous flux periodicity in proximitised quantum spin Hall constrictions. {\em New J. Phys.} {\bf 2022}, {\em 24}, 053017.

\bibitem{fT}
    Virtanen, P.; Recher, P. Signatures of Rashba spin-orbit interaction in the superconducting proximity effect in helical Luttinger liquids. {\em Phys. Rev. B} {\bf 2012}, {\em 85}, 035310.    

\bibitem{fC}
    Baxevanis, B.; Ostroukh, V.P.; Beenakker, C.W.J. Even-odd flux quanta effect in the Fraunhofer oscillations of an edge-channel Josephson junction. {\em Phys. Rev. B} {\bf 2015}, {\em 91}, 041409(R).

\bibitem{klinovaja}
    Legg, H.F.; Laubscher, K.; Loss, D.; Klinovaja, J. Parity protected superconducting diode effect in topological Josephson junctions. {\em arXiv} {\bf 2023}, arXiv:2301.13740.
    
\end{thebibliography}
\end{document}